\documentclass[prd,aps,twocolumn,a4paper,floatfix]{revtex4}
\def\p{\partial}
\def\Lie{{\cal L}}
                     
\def\half{\frac{1}{2}}
\usepackage{graphicx,psfrag}
\usepackage{mathrsfs}
\usepackage{amsmath,amsfonts,amssymb}

\begin{document}


\title{
  Constraint violation in free evolution schemes: \\
  comparing BSSNOK with a conformal decomposition of Z4
}

\author{Sebastiano \surname{Bernuzzi} and David \surname{Hilditch}}
\affiliation{Theoretical Physics Institute, University of 
  Jena, 07743 Jena, Germany}

\date{\today}

\begin{abstract}
We compare numerical evolutions performed with the BSSNOK 
formulation and a conformal decomposition of a Z4-like 
formulation of General Relativity. The 
important difference between the two formulations is that the 
Z4 formulation has a propagating Hamiltonian constraint, whereas 
BSSNOK has a zero-speed characteristic variable in the constraint 
subsystem. In spherical symmetry we evolve both puncture and 
neutron star initial data. We demonstrate that the propagating 
nature of the Z4 constraints leads to results that compare 
favorably with BSSNOK evolutions, especially when matter is 
present in the spacetime. From the point of view of 
implementation the new system is a simple modification 
of BSSNOK.
\end{abstract}

 
\maketitle


\section{Introduction}
\label{sec:Introduction}


Numerical evolutions of the Einstein equations are complicated 
by the constraints of the system. At the continuum level the 
constraint subsystem of a given formulation of General Relativity 
(GR) must close; if the constraints are satisfied in one 
time-slice, they will remain so. There are two possible numerical 
approaches to the constraints of the system. The first is to 
perform a so-called constrained evolution, in which the 
constraints are explicitly resolved at every point in time. There 
has been some progress in this direction in 3-D numerical relativity 
since the work of Bonazzola et al.~\cite{Bonazzola:2003dm} (see also
\cite{CorderoCarrion:2008cx,CorderoCarrion:2008nf} for mathematical 
analysis and recent developments), but the fact that the 
constraint equations are elliptic in character is a
limitation from the point of view of computational cost. The 
second, more common approach is to perform a free evolution. 
For free evolution the constraints are explicitly solved only 
in the initial time-slice. Throughout an evolution the constraints 
will be violated by numerical error, but this error should converge 
away with resolution. Computational resources are however 
always limited and in some situations a large constraint 
violation can be responsible for a numerical evolution to fail. 
Empirically spacetimes which contain matter are especially likely 
to suffer from large constraint violation. 

A major research topic has been the development of methods to 
minimise constraint violation at finite resolution. An obvious 
example is the PDEs and numerical analysis of the constraint 
subsystem of a given formulation. Formulations with a constraint 
system whose characteristic speeds are non-zero are preferable, 
since one may hope that constraint violations propagate out of the 
numerical domain as opposed to sitting on the grid and growing. 
A related issue is the construction of constraint preserving 
boundary conditions, which prevent large constraint violations 
from propagating in from the outer boundary of the computational 
domain. Here we consider the effect of propagating constraints 
in combination with a damping scheme. 

The idea behind the constraint damping method is to identify places 
in the main system to subtract constraints, such that the implied 
constraint subsystem picks up terms that force a reduction in the 
size of any violation.  In fact the constraint damping scheme of 
Gundlach et al.~\cite{Gundlach:2005eh} was used in the first successful 
numerical simulation of the binary black hole merger to dramatic 
effect. Essentially the same scheme is now used in the state of 
the art pseudo-spectral code of the Caltech-Cornell collaboration, 
see for example~\cite{Scheel:2008rj} and in the finite difference 
codes~\cite{Palenzuela:2009yr,Szilagyi:2006qy}. All of these 
successful uses of constraint damping have been for the 
generalized harmonic formulation of the Einstein equations, 
which implicitly relies on gauge conditions that are in the 
principal part harmonic.

On the other hand, many groups have been using the BSSNOK
formulation \cite{Nakamura:1987,Shibata:1995we,Baumgarte:1998te}
(hereafter BSSN for brevity) with the l+log and 
$\tilde{\Gamma}$-driver gauge conditions to perform simulations of 
both matter and vacuum space-times, following the approach 
of~\cite{Baker:2005vv,Campanelli:2005dd}. Unfortunately, there 
has not been a systematic comparison of the constraint violation 
between the different approaches, but every piece of evidence 
indicates that the simulations give comparable, 
compatible results. The differences between the generalized harmonic 
and puncture approaches prevent the two methods from being 
straightforwardly interchanged. For example it is not possible to 
evolve the standard puncture initial data with the generalized 
harmonic approach. BSSN on the other hand lacks a constraint 
damping scheme. In this work we demonstrate that it is possible 
to evolve spacetimes of interest with a conformal 
decomposition of the Z4~\cite{Bona:2003fj,Bona:2004yp} formulation,
which we denote Z4c (conformal Z4). Furthermore we show that the 
constraint damping scheme for Z4 can be applied effectively and 
straightforwardly to the modified system. We find the new system 
particularly efficient in the evolution of non-vacuum spacetimes. 
From the implementation point of view, the new system is a simple 
adjustment of the BSSN equations.

In section~\ref{sec:Formulation} the Z4c system
is described in some detail. Section~\ref{sec:Hydro} contains a 
brief review of the equations of relativistic hydrodynamics. Our 
numerical results are contained in section~\ref{sec:Numerial_Results}. 
Finally we conclude in section~\ref{sec:Conclusion}.

Through the paper we use geometrical units $c=G=1$, numerical 
results are reported with $M_{\odot}=M_{\rm bh}=1$.

\section{The Z4c system}
\label{sec:Formulation}

\paragraph*{Formulation:} The (constraint damped) Z4 
formulation~\cite{Bona:2003fj,Bona:2004yp,Gundlach:2005eh}
takes the 4-dimensional Einstein equations and replaces them by
\begin{align}
R_{ab}+\nabla_aZ_b+\nabla_bZ_a =
8\pi (T_{ab}-\frac{1}{2}g_{ab}T)\nonumber\\
+\kappa_1 [ t_aZ_b + t_bZ_a - (1+\kappa_2)g_{ab}t_cZ^c ],\label{eq:Z4_1}
\end{align}
where $Z_a$ is a 4-vector of constraints and $t^a$ is a time-like
vector field. Solutions of Eq.~\eqref{eq:Z4_1} are also valid solutions
of the Einstein equations when the constraints $Z_a$ vanish.
From the PDEs point of view, the most important part of the constraint 
addition is that of the partial derivatives. 
The {\it constraint damping } terms with coefficients $\kappa_i$ 
are discussed in more detail in the following section. We $3+1$ 
decompose the system and discard non-damping non-principal 
modifications to the ADM equations. This has the undesirable 
effect of breaking the 4-covariance of the Z4 formulation, but on 
the other hand we will see that discarding the lower order terms 
allows the evolution equations to be written very similarly to 
BSSN. The time-evolution equations are
\begin{align}
\p_t\gamma_{ij}&= \Lie_\beta \gamma_{ij} - 2\alpha K_{ij},
\label{eq:Z4_ADM_1}\\
\p_tK_{ij}     &=  -D_iD_j\alpha+\alpha[R_{ij}-2K_{ik}K^k_j+K_{ij}K
+2\p_{(i}Z_{j)}]\nonumber\\
&+\Lie_\beta K_{ij}+ 4\pi\alpha[\gamma_{ij}
(S-\rho_{{\textrm {\tiny ADM}}})-2S_{ij}]\nonumber\\
& -\kappa_1(1+\kappa_2)\alpha\gamma_{ij}\Theta,\\
\p_t\Theta &= \alpha[\frac{1}{2} H + \p_kZ^k
-(2+\kappa_2)\kappa_1\Theta]+\beta^i\Theta_{,i},\label{eq:Theta_dot}\\
\p_tZ_{i}  &= \alpha M_i + \alpha \Theta_{,i}-\alpha\kappa_1Z_i
+\beta^jZ_{i,j},
\label{eq:Z4_ADM_2}
\end{align}
where $Z_i$ is just the spatial projection of $Z_a$ and $\Theta=-n_aZ^a$.
We stress that these equations of motion are not identical to those of 
Z4, although they differ only in non-principal terms. The Hamilonian and 
momentum constraints are 
\begin{align}
H  &\equiv R-K_{ij}K^{ij}+K^2 - 16 \pi \rho_{{\textrm {\tiny ADM}}}=0,\\
M^i&\equiv D_j(K^{ij}-\gamma^{ij}K) - 8 \pi S^i  =0.
\end{align}
We adopt the standard notation for the metric variables and decompose 
the stress-energy tensor as 
\begin{align}
\rho_{{\textrm {\tiny ADM}}} &=n_an_bT^{ab},\\
S_i&=    -\gamma_{ia}n_bT^{ab},\\
S_{ij}&=  \gamma_{ia}\gamma_{jb}T^{ab}.
\end{align}
To write the system as similarly as possible as the BSSN formulation 
we now conformally decompose the variables, or in other words make 
the change of variables 
\begin{align}
\tilde{\gamma}_{ij} = \gamma^{-\frac{1}{3}}\gamma_{ij}, 
&\qquad 
\chi = \gamma^{-\frac{1}{3}},
\\
\hat{K} = \gamma^{ij}K_{ij} - 2\Theta,
&\qquad
\tilde{A}_{ij}=\gamma^{-\frac{1}{3}}(K_{ij}-\frac{1}{3}\gamma_{ij}K),
\end{align}
and
\begin{align}
\tilde{\Gamma}^{i} &= 
 2 \tilde{\gamma}^{ij} Z_j + \tilde{\gamma}^{ij} 
\tilde{\gamma}^{kl}\tilde{\gamma}_{jk,l},
\end{align}
The evolution equations are then
\begin{align}
\p_t \chi &= \frac{2}{3}\chi[\alpha(\hat{K}+2\Theta) - D_i\beta^i ]
\label{eq:Z4_decomp_first},\\
\p_t \tilde{\gamma}_{ij} &= -2\alpha\tilde{A}_{ij}+\beta^k
\tilde{\gamma}_{ij,k}+\tilde{\gamma}_{ik}\beta^k_{,j}-\frac{2}{3}
\tilde{\gamma}_{ij}\beta^k_{,k},\\
\p_t \hat{K}    &= -D^iD_i\alpha + \alpha[\tilde{A}_{ij}\tilde{A}^{ij}
+\frac{1}{3}(\hat{K}+2\Theta)^2]\nonumber\\
&+4\pi\alpha[S+\rho_{{\textrm {\tiny ADM}}}]+\beta^iK_{,i}
+\alpha\kappa_1(1-\kappa_2)\Theta
\end{align}
the trace-free extrinsic curvature evolves with
\begin{align}
\p_t \tilde{A}_{ij} &= \chi[-D_iD_j\alpha
+\alpha (R_{ij}-8\pi S_{ij})]^{\textrm{tf}}\nonumber\\
& +\alpha[(\hat{K}+2\Theta)\tilde{A}_{ij} - 2\tilde{A}^k{}_i\tilde{A}_{kj}]
\nonumber\\
& + \beta^k\tilde{A}_{ij,k}+\tilde{A}_{ik}\beta^{k}{}_{,j}
-\frac{2}{3}\tilde{A}_{ij}
\beta^{k}{}_{,k} 
\end{align}
and finally we have
\begin{align}
\p_t \tilde{\Gamma}^{i} &= -2\tilde{A}^{ij}\alpha_{,j}+2\alpha
[\tilde{\Gamma}^i_{jk}\tilde{A}^{jk}-2\tilde{A}^{ij}\ln(\chi)_{,j}
\nonumber\\
&-\frac{2}{3}\tilde{\gamma}^{ij}(\hat{K}+2\Theta)_{,j}
-8\pi\tilde{\gamma}^{ij}S_j]+\tilde{\gamma}^{jk}\beta^i_{,jk}\nonumber\\
&
+\frac{1}{3}\tilde{\gamma}
^{ij}\beta^k_{,kj}+\beta^j\tilde{\Gamma}^i_{,j}
-\tilde{\Gamma}_{\textrm{d}}{}^j\beta^i_{,j}+\frac{2}{3}
\tilde{\Gamma}_{\textrm{d}}{}^i\beta^j_{,j},
\end{align}
where we write $\tilde{\gamma}^{jk}\tilde{\Gamma}^{i}{}_{jk}=
\tilde{\Gamma}_{\textrm{d}}{}^i$. The $\Theta$ variable evolves according 
to Eq.~\eqref{eq:Theta_dot} with the appropriate 
substitutions Eqns.~(\ref{eq:Conf_Constr_1}-\ref{eq:Conf_Constr_2}). 
As with BSSN we write, in the $\p_t\tilde{A}_{ij}$ equations 
\begin{align}
R_{ij} &= R^{\chi}{}_{ij} + \tilde{R}_{ij},\\
\tilde{R}^{\chi}{}_{ij} &=
\frac{1}{2\chi}\tilde{D}_i\tilde{D}_j\chi+\frac{1}{2\chi}
\tilde{\gamma}_{ij}\tilde{D}^l\tilde{D}_l\chi\nonumber\\
&-\frac{1}{4\chi^2}\tilde{D}_i\chi\tilde{D}_j\chi-\frac{3}{4\chi^2}
\tilde{\gamma}_{ij}\tilde{D}^l\chi\tilde{D}_l\chi,\\
\tilde{R}_{ij} &=
 - \frac{1}{2}\tilde{\gamma}^{lm}
\tilde{\gamma}_{ij,lm} +\tilde{\gamma}_{k(i|}\tilde{\Gamma}
^k_{|,j)}+\tilde{\Gamma}_{\textrm{d}}{}^k
\tilde{\Gamma}_{(ij)k}+\nonumber\\
&\tilde{\gamma}^{lm}\left(2\tilde{\Gamma}^k_
{l(i}\tilde{\Gamma}_{j)km}+\tilde{\Gamma}^k_{im}
\tilde{\Gamma}_{klj}\right).
\end{align}
The complete set of constraints are, in terms of the evolved 
variables, given by
\begin{align}
\Theta  &,  \qquad 
2 Z_i     = \tilde{\gamma}_{ij}\tilde{\Gamma}^{j}-
             \tilde{\gamma}^{jk}\tilde{\gamma}_{ij,k},
\label{eq:Conf_Constr_1}\\
H       &= R -\tilde{A}^{ij}\tilde{A}_{ij}+\frac{2}{3}
           (\hat{K}+2\Theta)^2-16\pi\rho_{{\textrm {\tiny ADM}}},\\
\tilde{M}^i &= \p_j\tilde{A}^{ij} + \tilde{\Gamma}^i{}_{jk}\tilde{A}^{jk}
             -\frac{2}{3}\tilde{\gamma}^{ij}\p_j(\hat{K}+2\Theta)\nonumber\\
            & -\frac{3}{2}\tilde{A}^{ij}(\log\chi)_{,j},\\
D           &\equiv\ln(\det\tilde{\gamma})=0,\qquad
T            \equiv \tilde{\gamma}^{ij}\tilde{A}_{ij}=0\label{eq:Conf_Constr_2}
\end{align}
The change of variables introduces the algebraic constraints
$D$ and $T$. The numerical implementation explicitly imposes 
these constraints, so the continuum system we evolve is equivalent 
to Eqns.~(\ref{eq:Z4_ADM_1}-\ref{eq:Z4_ADM_2}). One may obtain BSSN from 
Z4c by taking $\Theta\to 0$ 
in Eqns.~(\ref{eq:Z4_decomp_first}-\ref{eq:Conf_Constr_2}).

\paragraph*{Hyperbolicity:} 
Statements about the hyperbolicity of Z4c obviously 
follow directly from those about Z4 since the two formulations share the 
same principal part. For completeness we summarize the relevant results 
here. A first order in time, second order in space system is called strongly 
hyperbolic if it has a complete set of characteristic variables with real 
characteristic speeds. This definition is equivalent to the existence
of a strongly hyperbolic first order reduction of the 
system~\cite{Gundlach:2004ri,Gundlach:2005ta}. The Z4c system 
is strongly hyperbolic when coupled to the puncture gauge conditions
\begin{align}
\label{eq:lapse_1log}
\p_t\alpha   &= \beta^i\alpha_{,i}-\alpha^2\mu_L\hat{K},\\
\label{eq:beta_Gdriv}
\p_t\beta^i  &= \mu_S\tilde{\Gamma}^i-\eta \beta^i+\beta^j\beta^i{}_{,j}
\end{align}
in which we always choose $\mu_L=2/\alpha$ and  $\mu_S=1$. To construct 
the characteristic variables we perform a $2+1$ decomposition in space and 
construct the fully second order characteristic 
variables~\cite{Hilditch:2010wp}. 
Defining $\tilde{\mu}_S=\gamma^{\frac{1}{3}}\mu_S$ and 
\begin{align}
\p_0=\frac{1}{\alpha}(\p_t-\beta^i\p_i),
\end{align}
the scalar sector has characteristic variables 
\begin{align}
U_{s\pm \mu_L}&= \p_0\alpha\pm\sqrt{\mu_L}\alpha_{,s},\\
U_{s\pm 1}&= \p_0\gamma_{qq}\pm\gamma_{qq,s},\\
U_{s\pm 1}{}'&= \p_0\gamma_{ss}\pm\frac{4}{3}\gamma_{ss,s}
\pm\frac{1}{3}\gamma_{qq,s}-\frac{2}{\alpha^2\mu_L}\p_0\alpha\nonumber\\
&-\frac{2}{\alpha}\beta_{s,s}\mp\frac{2}{\alpha^2\tilde{\mu}_S}\beta_{s,s},\\
U_{s\pm \frac{4}{3}\tilde{\mu}_S}&=
\p_0\beta_{s}\pm \frac{\sqrt{3}}{2}\sqrt{\tilde{\mu}_S}\beta_{s,s}\nonumber\\
&-\frac{\alpha\tilde{\mu}_S}{6(1-\tilde{\mu}_S)}
\big(\gamma_{ss,s}\mp\frac{\sqrt{3}}{2}
\sqrt{\tilde{\mu}_S}\p_0\gamma_{ss}\big)\nonumber\\
&-\frac{\alpha\tilde{\mu}_S}{6(1-\tilde{\mu}_S)}
\big(\gamma_{qq,s}\mp\frac{\sqrt{3}}{2}
\sqrt{\tilde{\mu}_S}\p_0\gamma_{qq}\big)\nonumber\\
&-\hat{\mu}(\alpha_{,s}-\sqrt{3\tilde{\mu}_S}
(1+\frac{1-\mu_L}{3-4\tilde{\mu}_S})\p_0\alpha)
\end{align}
with speeds $\pm(\sqrt{\mu_L},1,1,2\sqrt{\tilde{\mu}_S/3})$ respectively, where
\begin{align}
\hat{\mu}&=\frac{\tilde{\mu}_S(3-4\tilde{\mu}_S)}{(3\mu_L-4\tilde{\mu}_S)
(1-\tilde{\mu}_S)}.
\end{align}
The vector sector has characteristic variables
\begin{align}
U_{A\pm\sqrt{\tilde{\mu}_S}}&= \p_0\beta_A \pm \sqrt{\tilde{\mu}_S}\beta_{A,s},\\
U_{A\pm 1}&= \p_0\gamma_{sA}\pm\gamma_{sA,s}
-\frac{1}{\tilde{\mu}_S\alpha}(\tilde{\mu}_S\beta_{A,s}\pm\p_0\beta_A),
\end{align}
with speeds $\pm(\sqrt{\tilde{\mu}_S},1)$. Finally the tensor sector has 
characteristic variables
\begin{align}
U_{AB\pm 1}&= \p_0\gamma_{AB} \pm \gamma_{AB,s},
\end{align}
with speeds $\pm 1$. Note that the choice $\tilde{\mu}_S=1$ renders some 
of the characteristic variables singular. In that special case 
one should analyse the system independently; it is again strongly
hyperbolic, but we omit the characteristic variables. Strong 
hyperbolicity is enough to guarantee well-posedness of the Cauchy 
problem. For the initial boundary value problem, the stricter 
notion of symmetric hyperbolicity is desirable. We do not consider 
it here.

\paragraph*{Constraint subsystem:} The principal part of the 
Z4c constraint subsystem is given by
\begin{align} 
\p_0\Theta&\simeq\frac{1}{2}H+\p_iZ^i,\\
\p_0Z_i&\simeq M_i+\p_i\Theta,\\
\p_0H&\simeq -2\p_iM^i,\\
\p_0M_i&\simeq -\frac{1}{2}\p_iH+\p^j\p_jZ_i-\p_i\p^jZ_j.
\end{align}
Each of the variables satisfies a fully second order wave equation. 
The characteristic variables are simply
\begin{align}
\p_0\Theta\pm\Theta_{,s},&\qquad \p_0Z_i\pm Z_{i,s},\\
\p_0H  \pm H_{,s},       &\qquad \p_0M_i\pm M_{i,s},
\end{align} 
each with speeds $\pm 1$. In the numerical experiments that 
follow the crucial property is the propagating nature of the 
Z4c subsystem. The constraint damping scheme adds 
another level of complexity. However, if we linearize the 
system around flat-space we trivially recover the Z4 constraint
system; the analysis of~\cite{Gundlach:2005eh} holds 
immediately. All but constant frequency constraint violating 
modes are damped.

\paragraph*{Comparison with the BSSN constraint system:} 
For a complete discussion of the BSSN formulation we 
suggest~\cite{Baumgarte:2002jm,Gourgoulhon:2007ue,Alcubierre:2008}. Here we focus only 
on the BSSN constraint subsystem in order to compare with that 
of the Z4c system. Assuming that the algebraic 
constraints are satisfied, the principal part of the constraint 
system is given by
\begin{align}
\p_0Z_i&\simeq M_i,\qquad
\p_0H\simeq -2\p_iM^i,\\
\p_0M_i&\simeq \frac{1}{6}\p_iH+\p^j\p_jZ_i+\frac{1}{3}\p_i\p^jZ_j.
\end{align}
The characteristic variables are 
\begin{align}  
C_{s\pm 1}&=Z_{s,s}+\frac{1}{8}H \pm \frac{3}{4}M_s,\\
C_{s 0}&= \frac{1}{2}H + Z_{s,s}.
\end{align}
with speeds $(\pm 1,0)$ in the scalar sector and 
\begin{align}
C_{A\pm 1}&= Z_A \pm M_A,
\end{align}
with speeds $\pm 1$ in the vector sector. 

There are two important differences between the two constraint subsystems. 
The first is that BSSN does not accept a natural constraint 
damping scheme on every constraint. The second is the 0-speed 
characteristic variable in the BSSN system. In free-evolution
numerical applications, some violation of the constraints 
is inevitable. Since the numerical scheme will inherit 
properties from the continuum system, one expects errors in 
the BSSN Hamiltonian constraint to sit on the numerical grid 
and, potentially, grow. On the other hand one 
should expect Z4c Hamiltonian constraint violations to propagate.
In combination with suitable boundary conditions and the 
damping scheme we expect that Z4 Hamiltonian constraint 
violations will be easier to control than those of BSSN. 
The conclusions obtained from the above linear PDEs analysis may 
not carry over to fully non-linear evolutions. Numerical 
evolutions are required to test their validity.

\section{Relativistic hydrodynamics}
\label{sec:Hydro}

We assume that the matter is described by a perfect fluid 
stress-energy tensor:
\begin{align}
  \label{eq:tmunu}
  T_{ab}=\rho h u_a u_b + p g_{ab} \ ,
\end{align}
where $\rho$ is the rest-mass density,
$\epsilon$ is the specific internal energy,
$h\equiv1+\epsilon + p/\rho$ is the specific enthalpy,
$p$ is the pressure, 
and $u^a$ is the 4-velocity ($u^a u_a=-1$) of the fluid.
The total energy density is given by
$\varepsilon=\rho(1+\epsilon)$.

The General Relativistic HydroDynamics equations for the perfect 
fluid matter (\emph{ideal} GRHD) are the local conservation law for 
the energy-momentum tensor, the conservation law for the baryon 
number and the Equation of State (EoS) of the fluid:
\begin{align}
  \nabla_a T^{ab} & = 0, \label{eq:divT}\\
  \nabla_a \left( \rho u^a \right) & = 0, \label{eq:rhoua}\\
  P(\rho,\epsilon) &= p  \ .
\end{align}
Following \cite{Banyuls:1997} we rewrite Eqns.~(\ref{eq:divT}-\ref{eq:rhoua}) 
in first-order, flux-conservative form: 
\begin{equation}
  \label{eq:hydro_cons_form}
  \partial_t \vec{q} + \partial_i \vec{f}^{(i)} (\vec{q}) = \vec{s}(\vec{q}) \; ,
\end{equation}
by introducing the \emph{conservative} variables:
\begin{align}
  \vec{q} &= \sqrt{\gamma}\{ \, D, \, S_k, \, \tau \, \},
\end{align} 
where 
\begin{align}
  D &= W\rho, \\
  S_k &= W^2 \rho h  v_k, \\
  \tau &= \left(W^2 \rho h - p\right) - D \ .
\end{align}
The simple physical interpretation of these variables is that they represent 
the rest-mass density ($D$), the momentum density ($S_k$) and an internal
energy ($\tau=\rho_{\rm ADM}-D$) as viewed by Eulerian observers.
Above $v^i$ is the fluid velocity measured by the Eulerian observer:
\begin{equation}
  v^i = \frac{u^i}{W} + \frac{\beta^i}{\alpha} = 
  \frac{1}{\alpha}\left( \frac{u^i}{u^0}+ \beta^i \right) \ ,
\end{equation}
and $W$ is the Lorentz factor between the fluid frame and the Eulerian observer, 
$W=1/\sqrt{1-v^2}$, with $v^2=\gamma_{ij}v^i v^j$.
The fluxes in Eq.~(\ref{eq:hydro_cons_form}) are:
\begin{align}
  \label{eq:hydro_fluxes}
  \vec{f}^{(i)} &= \sqrt{-g}\left\{ D\left(v^i-\frac{\beta^i}{\alpha}\right), \right. \\ 
  & S_k\left(v^i-\frac{\beta^i}{\alpha}\right) + p\delta_k^i, \nonumber \\
  & \left. \tau\left(v^i-\frac{\beta^i}{\alpha}\right)+pv^i \right\} \nonumber
\end{align}
while the source terms are:
\begin{align}
  \vec{s} &= \sqrt{-g}\left\{0, \right.  
  \label{eq:hydro_sources}\\
  & T^{ab}\left( \partial_a g_{a k} - \Gamma^\delta_{ab}g_{\delta k}\right) , \nonumber\\ 
  & \left. \alpha\left( T^{a 0}\partial_a \ln \alpha - 
      T^{ab} \Gamma^0_{ab} \right) \right\}  \nonumber \\
  %
  %
  & = \sqrt{-g}\left\{0, \right. 
  \label{eq:hydro_sources_sder}\\
  & T^{00}\left (\half \beta^i\beta^j \partial_k \gamma_{ij} - \alpha \partial_k \alpha \right)  + \nonumber \\ 
  & \qquad + T^{0i}\beta^j\partial_k \gamma_{ij}  + T^0_i\partial_k\beta^i + \half T^{ij}\partial_k \gamma_{ij} ,\nonumber \\
  & \left. T^{00}\left( \beta^i\beta^j K_{ij} - \beta^i\partial_i \alpha\right)
    + T^{0i}\left( 2\beta^j K_{ij} - \partial_i \alpha \right)  +
    T^{ij}K_{ij} \right\} \nonumber
\end{align}
Above $g\equiv \det g_{ab}=-\alpha^2\gamma$, 
$\gamma\equiv \det \gamma_{ij}$ and 
Eq.~(\ref{eq:hydro_sources_sder}) can be obtained using Eq.~(\ref{eq:Z4_ADM_1}) in a way to eliminate time derivatives.
Note that both the fluxes and the source terms depend also on the
\emph{primitives} variables 
$\vec{w}=\{p, \rho, \epsilon, v^i \}$, and the source terms 
do not depend on derivatives of $T_{ab}$. 
The system in Eq.~(\ref{eq:hydro_cons_form}) is strongly hyperbolic 
provided that the EoS is causal (the sound speed is less than the speed 
of light)~\cite{Banyuls:1997}.

\section{Numerical results}
\label{sec:Numerial_Results}

We consider the following set of tests in spherical symmetry:
\begin{description}

\item[Flat spacetime] Evolution of constraint violating initial data. This
  test shows the basic mechanism of constraint progation and damping at work;

\item[Puncture spacetime] Evolution of puncture initial data. Z4c can be
  successfully used for the simulation of black-hole spacetimes with 
  the puncture gauge;

\item[Stable star] Evolution of a stable compact star. Constraint propagation
  and damping are particularly useful in the simulation of non-vacuum
  spacetimes in which zero speed modes related to the Hamiltonian constraint
  sit on the grid and grow during the evolution;
 
\item[Migrating star] Evolution of an unstable star which migrates towards a stable
  one. The main features of Z4c are demonstrated in this nonlinear, very 
  dynamical scenario; 

\item[Collapsing star] Evolution of an unstable star which collapses to a
  black hole. This test involves all of the difficulties of a simulation 
  in numerical relativity: matter, nonlinear dynamics, formation of a singularity 
  and black hole evolution.

\end{description}
The tests are also performed with BSSN for comparison. We stress that the differences 
between the results are convergent features, and with sufficiently high resolution 
they can be made arbitrarily small. In the presentation of the results we 
focus on the differences in the behavior of the Hamiltonian constraint. There is 
not a significant difference between the dynamics of the momentum and $Z_i$ 
constraints of the two formulations. In Z4c evolutions the $\Theta$ 
constraint violation is typically of the same order as that of the Hamiltonian 
constraint.

In each simulation we take for simplicity the constraint damping parameters 
$\kappa_2=0$ and $\kappa_1\equiv k=\{0,0.02\}$, for a Z4c undamped/damped respectively. 
There is no particular justification for the value $k=0.02$. The value was found to 
be reasonable after the first numerical experiments. More detailed discussion follows. 
A systematic study of the $\kappa$'s is left for future work.

The gauge choice is given by Eq.~(\ref{eq:lapse_1log}) and Eq.~(\ref{eq:beta_Gdriv}), 
one flavour of the popular puncture gauge~\cite{vanMeter:2006vi}. In 
Eq.~(\ref{eq:beta_Gdriv}) we use $\eta=2/M_{\rm ADM}$; choices more popular for matter 
evolutions, e.g. $\eta=0.3/M_{\rm ADM}$ \cite{Montero:2008yx}, were tested in some cases, 
but without significant differences. 

The equations solved in these tests are obtained by a \emph{faithful}
spherical reduction of the 3D Cartesian equations presented in 
Sec.~\ref{sec:Formulation}. Details are given in 
Appendix~\ref{sec:Sphere_Reduction}. Standard numerical methods based on 
finite-differences were used to solve them. They are described in 
Appendix~\ref{sec:Numerics}. While in 1D we may afford much higher 
resolutions than those of 3D simulations, we restrict to values reasonable 
for 3D mesh-refinement-parallel codes (see for instance Tab.~\ref{tab:settings}).
All the figures refer to the highest resolution simulations.

Initial stellar models are built using the polytropic EoS (see
Eq.~(\ref{eq:eos_poly})) with adiabatic index $\Gamma=2$ and polytropic
constant $K=100$ widely used in literature, e.g. Ref.~\cite{Font:2001ew}. 
The stable star model used in simulations in Sec.~\ref{sbsec:star} 
has central density $\rho_c=1.280\times10^{-3}$,
gravitational mass $M=1.400$ and circumferential radius $R=9.586$
(isotropic coordinate radius $r_R=8.126$).  
The unstable model used in Sections~\ref{sbsec:migr}
 and~\ref{sbsec:collapse} has central density $\rho_c=7.993\times10^{-3}$,
gravitational mass $M=1.448$ 
and circumferential radius $R=5.838$ (isotropic coordinate radius
$r_R=4.268$). 
The EoS employed during the evolution is the ideal gas:
\begin{equation}
  \label{eq:eos_ideal}
  P(\rho,\epsilon) = (\Gamma-1)\rho\epsilon \ .
\end{equation}

\subsection{Flat-space test}

\begin{figure}[t]
  \begin{center}
    \includegraphics[width=0.5\textwidth]{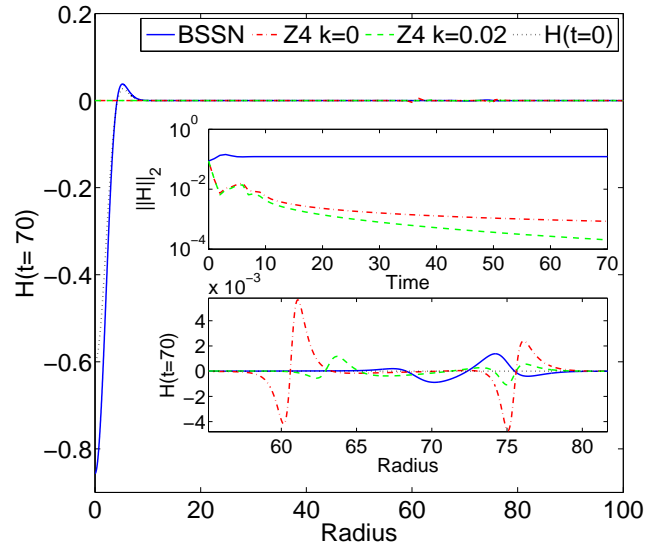}\\
    \caption{ \label{fig:H_flat} Hamiltonian constraint violation in the flat
      spacetime test for BSSN and Z4c. The main panel shows the 
      violation in space at the initial and final time of the 
      simulation. The lower inset focuses on large radii at the 
      final time, where a pulse is propagating outwards (see text
      for discussion). The upper inset shows the 2-norm of the 
      constraint in time.}
  \end{center}
\end{figure}

We begin by evolving a large constraint violating perturbation on
flat-space in standard spherical coordinates.
The violation is constructed by adding
\begin{align}
\delta\chi&= \exp( - r^2/10)
\end{align}
to the $\chi$ variable. According to section~\ref{sec:Formulation} 
we should find that the Z4c Hamiltonian constraint will 
propagate away from the origin, whereas the BSSN Hamiltonian 
constraint should split into two components, one stationary 
and one propagating. Fig.~(\ref{fig:H_flat}) demonstrates exactly 
the expected behavior. The main panel shows the Hamiltonian violation in
space at the final simulated time for the different formulations. In the 
BSSN evolution the violation near the origin grows (cf. the dotted line, which 
is $t=0$), while for the Z4c formulation it has propagated away. The 
propagation is clear from the lower inset, which shows the violation at 
large radius; the initial violation propagates out almost completely in 
a form of a pulse in case of Z4c, and only partially in case of BSSN (the 
pulse is smaller). Note in addition that the damping mechanism is working 
effectively in Z4c with $k=0.02$, for which the pulse become progressively 
smaller in space. Finally in the upper inset the 2-norms are presented. 
The global behavior of the constraint violation is better for Z4c as 
time advances.

Note that it is possible to see \emph{over-damping} effects in the
Z4c simulations. In fact we experimented with different damping values
and for example with  $k=2$ we found a large constraint-violating ``tail''
left behind the outgoing pulse near the origin. 

The effect of constraint damping has been demonstrated mathematically 
around flat-space for non-constant in space violations with a plane
wave ansatz. It is not entirely 
clear when that analysis will cease to determine the effectiveness of 
the constraint damping scheme.

\subsection{Puncture}

\begin{figure}[t]
  \begin{center}
    \includegraphics[width=0.5\textwidth]{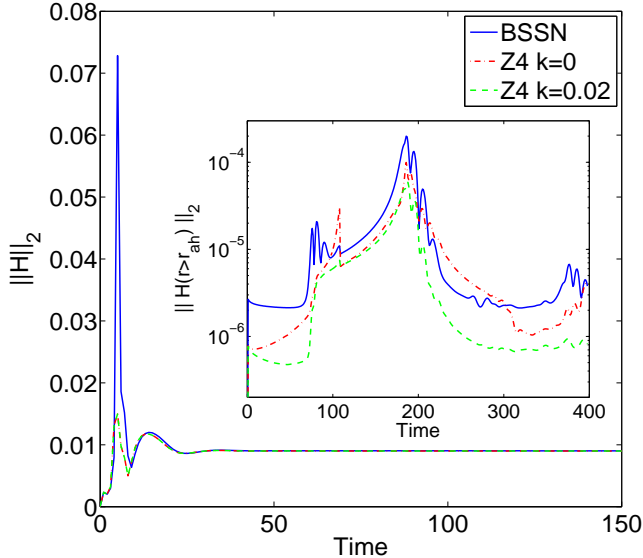}\\
    \caption{ \label{fig:n2H_punc} Hamiltonian constraint violation in the
      puncture test for BSSN and Z4c. The 2-norm of the constraint 
      is plotted in time. The inset shows the relative difference 
      between the norms of the two simulations.}
  \end{center}
\end{figure}

We evolve a single stationary puncture with a precollapsed lapse and
initially vanishing shift~\cite{Brugmann:2008zz}. 
Non-trivial evolution occurs due to the gauge adjustment near the puncture. We find 
little difference in the evolution with the two formulations, 
as well as in violation of the Hamiltonian constraint.

Gauge dynamics dominate the evolution until roughly $t\sim25$. 
After that the simulations settle at the stationary $1+\log$ trumpet 
solution~\cite{Hannam:2006vv}. Direct comparison with 3D BSSN 
simulations of the BAM code~\cite{Brugmann:2008zz} shows near 
identical evolution of the metric fields. 
 
The constraint violation results are summarized in Fig.~(\ref{fig:n2H_punc})
which shows the 2-norm of the Hamiltonian constraint during the evolution.
The initial gauge-transient and the stationary phase are clearly recognizable.
During the evolution the norm of Z4c is generally slightly better,
see for example the inset in Fig.~(\ref{fig:n2H_punc}) in which 
the 2-norm computed outside of the horizon is plotted. 

Looking at the spatial violation of the Hamiltonian constraint, there is
large violation near the puncture, where the numerical solutions 
are almost identical. The violation is unsurprising since we are 
finite differencing across an irregular solution. Constraint damping 
appears to have no effect in a neighborhood of the puncture. Away 
from the puncture we find that there are small differences in 
the Hamiltonian constraint violation. All of the simulations 
in Fig.~(\ref{fig:n2H_punc}) were performed with Sommerfeld 
boundary conditions. Incoming constraint violation from the 
boundary can be seen at $t=200$ in the inset of 
Fig.~(\ref{fig:n2H_punc}).

\subsection{Stable star}
\label{sbsec:star}

\begin{figure}[t]
  \begin{center}
    \includegraphics[width=0.5\textwidth]{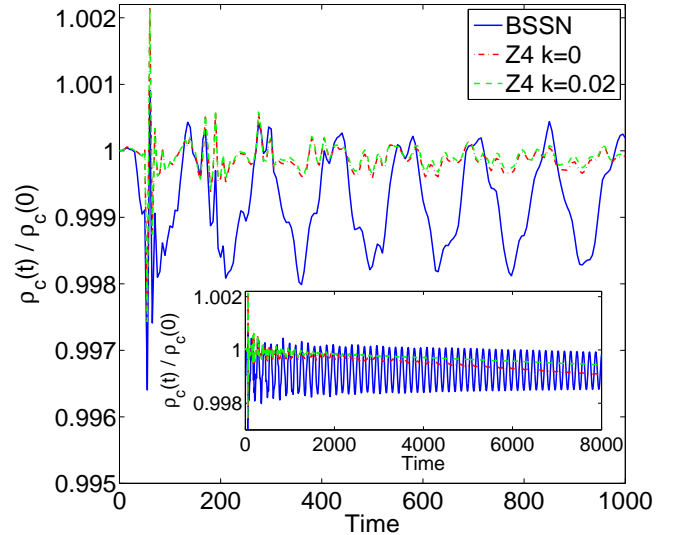}\\
    \caption{ \label{fig:rhoc_star} Oscillations of the central energy density
      in time in the star test for BSSN and Z4c. The inset shows the same for
      the whole simulation.} 
  \end{center}
\end{figure}

\begin{figure}[t]
  \begin{center}
    \includegraphics[width=0.5\textwidth]{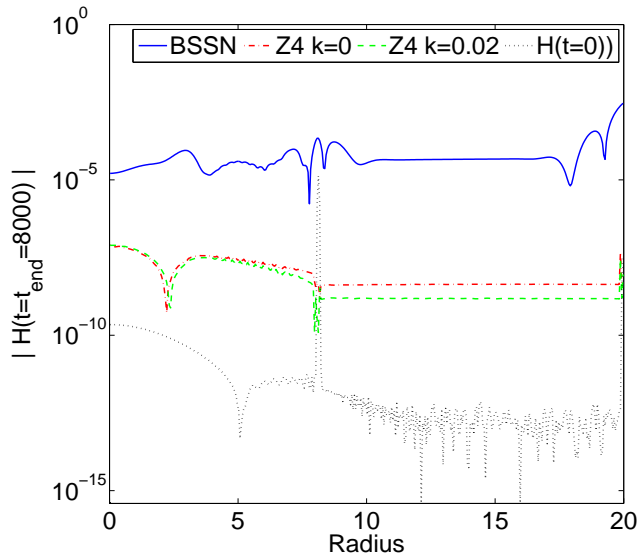}\\
    \caption{ \label{fig:H_star} Hamiltonian constraint violation in the
      stable star test for BSSN and Z4c. The zero speed mode of 
      the BSSN constraint subsystem causes a slow growth; such growth 
      does not occur in the Z4c evolution. The Z4c constraint 
      damping scheme reduces the size of the violation.
 }
  \end{center}
\end{figure}

\begin{figure}[t]
  \begin{center}
    \includegraphics[width=0.5\textwidth]{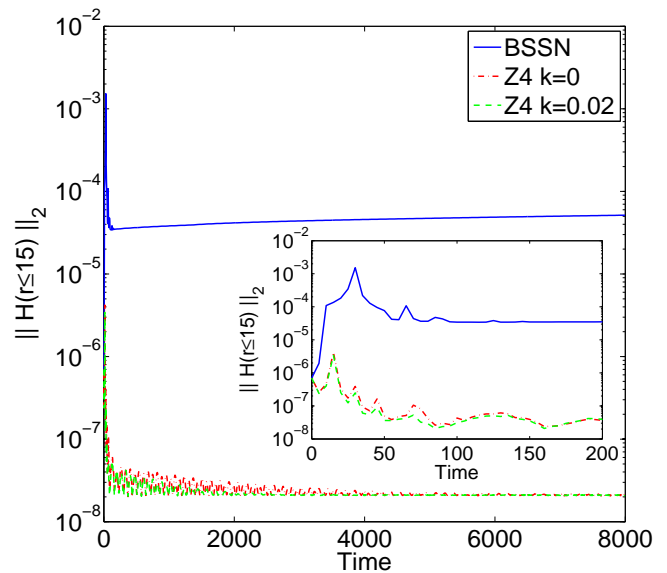}\\
    \caption{ \label{fig:n2H_star} Hamiltonian constraint violation in the
      stable star test. The 2-norm of the constraint up to $r=15$ is plotted in
      time. The inset shows a zoom at early times.} 
  \end{center}
\end{figure}

We evolve a stable spherical star to $t_{\textrm{end}}=8000$ (around
$40$ ms). 
As is standard in such simulations, truncation error causes the star 
to oscillate at its proper radial frequency. At late times oscillations are eventually 
damped by the numerical viscosity and by the interaction with the artificial 
atmosphere; a small linear drift of the mean value is usually also
observed,~\cite{Font:1999wh,Font:2001ew,Baiotti:2004wn}. 
The numerical error is constraint violating. 
In this context we expect our linear PDEs analysis to be a good guide 
to the behavior of the non-linear system, since the unperturbed background solution 
is static and the numerical error is a small perturbation. 

Figure~(\ref{fig:rhoc_star}) shows the radial oscillations of the central
rest-mass density in time. At the same resolution the amplitude of the
oscillations in the Z4c evolutions are significantly reduced with respect to BSSN. 
Additionally at late times they are completely suppressed, as it is clear from
the inset, and a drift, which improves in the constrained damped Z4c, can also 
be seen. The frequencies of the radial mode and its overtones agree in 
both BSSN and Z4c with the correct values, see e.g.~\cite{Baiotti:2008nf}
for both perturbative and 3D numerical results. 

The dynamics of the Hamiltonian constraint in the BSSN and Z4c evolutions 
are similar at early times. After the interpolation from our spectral 
initial data grid to the evolution grid, finite difference derivatives are 
used to evaluate the Hamiltonian constraint. This produces a relatively large 
violation in the initial data at the surface of the star. The violation in 
space at the end of the simulation is plotted in Fig.~(\ref{fig:H_star}). The 
dotted line in the figure is the initial violation. In both systems we find that 
this violation propagates away from the surface of the star. However in the BSSN 
evolution, the propagation leaves behind a stationary violation that grows 
linearly in time. This accumulation affect in BSSN is demonstrated in
Figs.~(\ref{fig:H_star}-\ref{fig:n2H_star}).

At the outer boundary a large violation of the constraint is clearly visible in 
the BSSN evolution. It is caused by the Sommerfeld boundary conditions used with 
BSSN. This violation is non-convergent but it is usually mitigated by
evolving with outer boundary further out, where the Sommerfeld conditions 
fare much better. The zero speed mode of BSSN plays the important role of keeping 
this violation at the boundary, thus minimizing their effect on the dynamics. 
On the other hand we find that for Z4c with Sommerfeld conditions the 
constraint violation from the outer boundary propagates inside the numerical 
domain and further perturbs the star, changing for instance the value of the 
central density and exciting very large amplitude oscillations. The issue is 
completely solved by the use of maximally dissipative boundary conditions (see
Appendix~\ref{sec:Numerics}), which were used in all the matter simulations
and in all of the presented figures. One possibility for not finding the same 
problem in our puncture evolutions is that in that case the constraint violation 
from the boundary is typically much smaller than the violation at the puncture and so 
does not significantly alter the dynamics. Also in the black hole evolution we 
evolve with a much larger outer boundary.

In Fig.~(\ref{fig:n2H_star}) we plot the 2-norm of the Hamiltonian constraint in 
time. The linear growth of the BSSN Hamiltonian constraint is visible when the
norm is computed up to $r\leq15$ instead of the full grid. The norm on the full 
grid is in fact dominated by the violation caused by the Sommerfeld boundary 
conditions (see also again Fig.~(\ref{fig:H_star})). The reason is that the 
violation inside the star converges. At lower resolutions 
however the linear growth of the Hamiltonian constraint is visible in the norm 
on the full grid. Note that although Fig.~(\ref{fig:n2H_star}) shows that the 
norm of the Hamiltonian constraint in the Z4c evolution drops  
below the initial value, the reason is the violation at the surface of the star 
in the initial data. Fig.~(\ref{fig:H_star}) shows that pointwise, at the end 
of the evolution the constraint violation has not dropped below the initial 
value except in a neighborhood of the surface of the star.

Our experiments indicate that the propagation 
of the constraints is more helpful in preventing growth than 
constraint damping.

\subsection{Migration}
\label{sbsec:migr}

\begin{figure}[t]
  \begin{center}
    \includegraphics[width=0.5\textwidth]{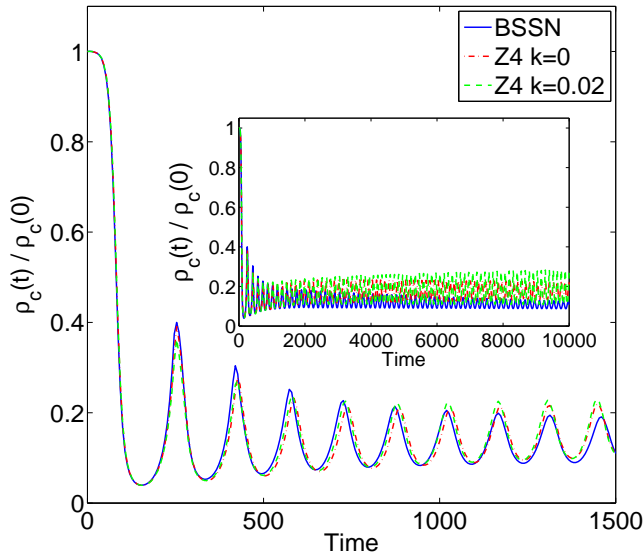}\\
    \caption{ \label{fig:rhoc_migr} Oscillations of the central energy density
      in time in the migration test for BSSN and Z4c.}
  \end{center}
\end{figure}

\begin{figure}[t]
  \begin{center}
    \includegraphics[width=0.5\textwidth]{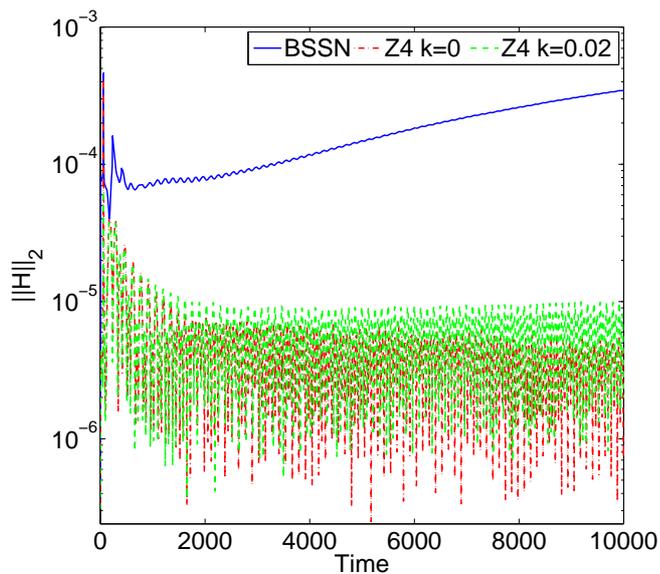}\\
    \caption{ \label{fig:n2H_migr} Hamiltonian constraint violation in the
      migration test. The 2-norm of the constraint is plotted in
      time.}
  \end{center}
\end{figure}

For the migration test we evolve initial data in hydrostatic 
equilibrium that is unstable against first order perturbations.
Truncation error causes the star to migrate to a model in the 
stable branch of the equilibrium configuration space. The energy 
difference between the two models is converted into kinetic 
energy and generates large nonlinear pulsations. With an ideal 
EoS the transition involves the formation of a shock wave 
between an inner core and an outer (lower density) mantle which 
dissipates kinetic energy into thermal energy. Furthermore a  
small amount of mass is expelled from the star when the shock 
reaches the surface. The dynamics are described in detail 
in~\cite{Font:2001ew,CorderoCarrion:2008nf}. This test is 
important because it probes the behavior of the two 
formulations in a genuinely non-trivial situation, 
where the applicability of the linear analysis is unclear.

Figure~(\ref{fig:rhoc_migr}) shows the central density during 
the simulation: the strong nonlinear oscillations are clearly 
visible and each is accompanied by a bounce of the core of the 
star which feeds the shock. At early times the dynamics are 
comparable with very small differences. At late times (see 
the inset) however we find some differences between both 
BSSN and Z4c both with and without the damping scheme. 
We believe the simulations at this point are not quantitatively 
reliable because of the lack of accuracy and convergence, so 
possible physical differences are not distinguishable from 
numerical errors. 

A second simulation carried with polytropic EoS (see Appendix~\ref{sec:Numerics},
Eq.~(\ref{eq:eos_poly})), which excludes a priori the presence of
shock heating~\footnote{Another difference when a one parameter EoS $P(\rho,\epsilon)=P(\rho)$
  is used is that the GRHD equation for $\tau$ is equivalent to the equation for $D$
  and thus need not be evolved.}, completely solves this issue. 
Even at late times the solutions are comparable 
and the results obtained with BSSN and Z4c with $k=0$ are basically the same.
Oscillations in this case are a little damped only by the numerical dissipation.
We furthermore observe that the data obtained with Z4c and $k=0.02$ are
affected by some additional damping.

The Hamiltonian violation behavior is qualitatively the same as 
in the stable star test. In Fig.~(\ref{fig:n2H_migr}) the 2-norm 
is plotted: the linear growth in the BSSN evolution is clearly 
visible (note that the boundary in this case is further out, so 
the Sommerfeld boundary condition contributes relatively little to 
the global violation) while the Z4c evolutions show a 
propagating and decreasing violation.
As opposed to the stable star case, at late times $t>2000$ 
the 2-norm for Z4c damped is larger than Z4c undamped, but 
this is again a numerical artifact due to the reasons discussed 
above. The simulations performed with the polytropic EoS result in
constraint violation similar to that in Fig.~(\ref{fig:n2H_star}).

\subsection{Collapse}
\label{sbsec:collapse}

\begin{figure}[t]
  \begin{center}
    \includegraphics[width=0.5\textwidth]{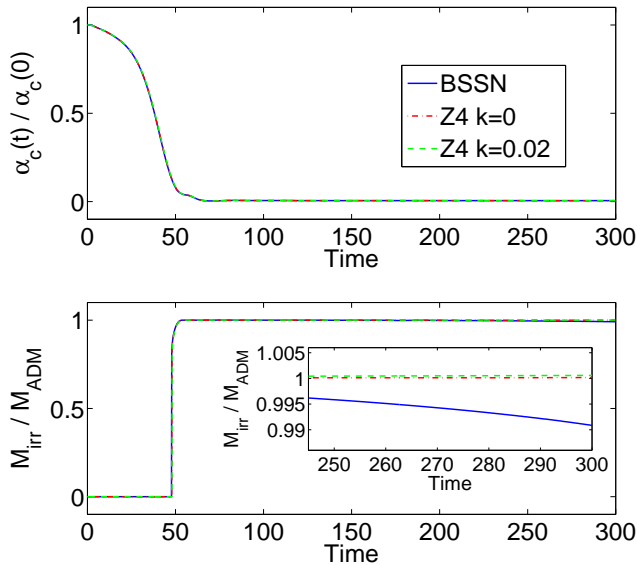}\\
    \caption{ \label{fig:Mirr_collapse} Collapse dynamics: 
      (upper panel) central value of the lapse and (bottom panel) 
      irreducible mass of the black hole 
      are plotted in in time in the collapse test for BSSN and Z4c. 
      The inset in the bottom panel shows a zoom of the irreducible 
      mass at late time.
    }
  \end{center}
\end{figure}

\begin{figure}[t]
  \begin{center}
    \includegraphics[width=0.5\textwidth]{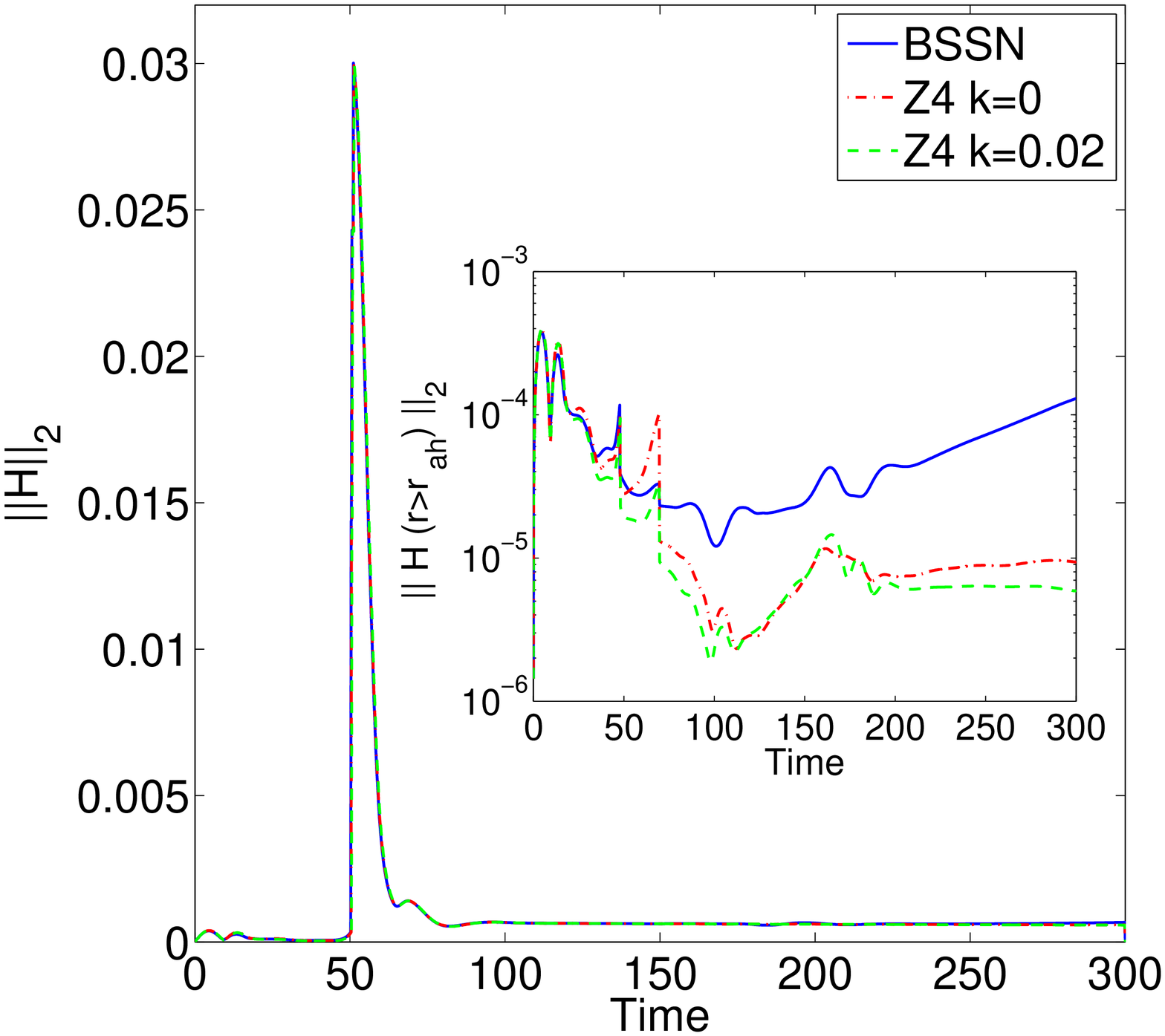}\\
    \caption{ \label{fig:n2H_collapse} Hamiltonian constraint violation in the
      collapse test. The 2-norm of the constraint is plotted in
      time. The inset shows the 2-norm computed only outside the apparent
      horizon; notice the log-scale of the y-axis.}
  \end{center}
\end{figure}

Our final test is the evolution of the unstable star, which  we perturb by a small 
(momentum constraint violating) function
\begin{equation}
\delta v^r = - 0.005 \sin(\frac{\pi r}{R}),
\end{equation}
to trigger collapse to a black hole. The solution space of the two formulations 
is not the same when the constraints are not satisfied. However we are able to 
follow the collapse with both BSSN and Z4c, and do not find significant 
differences in the dynamics of the two systems. This can be seen for instance 
in the upper panel of Fig.~(\ref{fig:Mirr_collapse}) in which we plot the 
central lapse in time. During the collapse the rest-mass is conserved
to within a $1\%$ error even in our lowest resolution simulation.

The Hamiltonian constraint violation of the formulations as the apparent 
horizon forms ($t\simeq 50$) is comparable Fig.~(\ref{fig:n2H_collapse}), but 
afterwards we find once again the linear growth in time of the BSSN 
Hamiltonian constraint. Finally the lower panel of 
Fig.~(\ref{fig:Mirr_collapse}) shows that after the collapse the irreducible 
mass of the black hole:
\begin{equation}
M_{\rm irr} = \sqrt{\frac{A_{\rm ah}}{16\pi}}
\end{equation}
where $A_{\rm ah}$ is the area evaluated from the apparent horizon,
drifts from the correct value by roughly $1\%$ in our BSSN evolution. At our lowest 
resolution (see Appendix~\ref{sec:Numerics}) this drift is a serious~($\simeq
25\%$ by $t=300$) problem, but it  
converges away at second order. On the other hand, the drift 
does not occur in any of the Z4c tests. Even at the lowest 
resolution the Z4c mass drift is below $0.1\%$. We mention 
that in Ref.~\cite{Montero:2008yx}, in which a different value 
of $\eta$ in the gauge condition is used, this effect is not 
seen.

\section{Conclusion}
\label{sec:Conclusion}

Despite their success in the evolution of binary black hole and 
neutron star systems, the solutions of the free-evolution schemes 
widely used in numerical relativity simulations have some constraint 
violation. Motivated by the observation that BSSN evolutions of matter 
spacetimes typically have much larger Hamiltonian than (the almost 
negligible) momentum constraint violation, we considered the PDE 
properties of the BSSN constraint subsystem. The constraint subsystem 
has a zero speed characteristic variable involving the Hamiltonian 
constraint, which we argue is the cause of the large violation. 
To demonstrate this with the aim of further improving the 
formulation for free-evolution schemes, we compare BSSN evolutions 
with those of a new conformal decomposition of the Z4 formulation. 
The latter formulation, referred as Z4c, has three desirable 
properties. Firstly the convenient choice of variables enables the 
evolution of puncture initial data. Secondly it inherits the 
propagating constraints of undecomposed system. Finally the 
constraint damping scheme described in~\cite{Gundlach:2005eh} can 
be attached to the new system.

In a large set of tests in spherical symmetry, performed 
with faithful spherical reductions of the two Cartesian systems, 
we have demonstrated that the BSSN constraint violation can indeed 
be avoided by using a system with propagating constraints. We 
focus on the differences in the Hamiltonian constraint violation 
because other variables do not show significant differences.
We find that the propagation of the constraints is more helpful 
than the constraint damping scheme in controlling the violation.
We also find that when using the Z4c formulation boundary 
conditions are typically more important than when using BSSN. 
In our tests we find that maximally dissipative conditions are 
always sufficient to avoid large constraint violation entering 
the grid from the outer boundary in Z4c evolutions.

Specifically we perform five tests, and focus on constraint 
violation. We always find qualitatively the same behavior 
(similar orders of magnitude) in the momentum constraints 
of the different systems. Therefore we focus on the behavior of the
Hamiltonian constraint. Firstly with a constraint 
violating perturbation on flat-space we find that the 
Z4c (BSSN) Hamiltonian constraint does (not) propagate, and 
that the Z4c constraint damping scheme works. Next we evolve 
a single puncture and find only small differences in the 
evolutions. When evolving either a stable or migrating star we 
find that at typical resolutions the Z4c Hamiltonian 
constraint violation is three or four orders of magnitude 
lower than that with BSSN. The migration test is especially 
important because it involves long-term nonlinear evolution. Our 
final test is the evolution of a star collapsing to a black hole. We find 
that the two formulations give very similar dynamics, which is 
not guaranteed because our initial data is constraint violating. 
We find that after the collapse the BSSN Hamiltonian constraint 
grows linearly in time, and that the irreducible mass of the 
black hole is not perfectly conserved at low resolutions. These 
problems are not present in the Z4c tests.

In the work presented here we have not tuned the constraint 
damping parameters, so an obvious step is to investigate the effect
of these parameters in more detail. However the most important 
follow up is to investigate whether or not similar results can
be reproduced in 3D, where there may be additional issues affecting
numerical stability. If 3D astrophysical simulations with Z4c are 
possible it will be interesting to see whether or not the improved 
Hamiltonian constraint behavior will effect the physics, in 
particular the extracted gravitational waves from binary 
systems or stellar collapse. 

\acknowledgements

It is a pleasure to thank Bernd Br\"ugmann, Roman Gold, 
Carsten Gundlach, Pedro Montero and Milton Ruiz for valuable
discussions and comments on the manuscript. We also wish to thank Luca
Del Zanna, Harry Dimmelmeier, Toni Font and Ian Hawke for helpful
advice on HRSC schemes. This work was supported in part by DFG grant  
SFB/Transregio~7 ``Gravitational Wave Astronomy''.

\appendix

\section{Spherical reduction of the equations}
\label{sec:Sphere_Reduction}

In spherical symmetry the line-element is:
\begin{equation}
  \label{eq:sphere:metric}
\textrm{d}s^2 = 
-\left(\alpha^2-\beta_r\beta^r\right)\textrm{d}t^2 + 
2\beta_r\textrm{d}r\textrm{d}t+\gamma_{rr}\textrm{d}r^2+
\gamma_{T}r^2 \textrm{d}\Omega^2 \ ,
\end{equation}
where $\beta_r = \beta^r/\gamma_{rr}$.
The determinant of the 3-metric can be written as:
\begin{equation}
  \label{eq:sphere:detg}
  \gamma = r^4 \sin^2 \theta \gamma^s \ ,
\end{equation}
with the definition $\gamma^s\equiv\gamma_{rr}\gamma^2_{T}$.
For the spherical reduction of the metric equations 
we follow~\cite{Garfinkle:2007yt}. An auxiliary flat derivative
whose connection vanishes in Cartesian coordinates is defined.
The standard formulation of the equations of motion is given in 
Cartesian coordinates. Thus the spherical reduction is made by 
replacing partial derivatives in constraint addition terms and 
non-tensorial variables with the auxiliary derivative, and simply 
rotating from Cartesian to spherical coordinates assuming two 
angular killing vectors. The resulting expressions are too
unwieldy to be presented here. 

The spherical reduction of the hydrodynamics equations of 
motion~(\ref{eq:hydro_cons_form}) is performed defining
conservative variables as:
\begin{equation}
  \vec{q} \equiv \sqrt{\gamma^s}\left\{D, S_r, \tau\right\} \ .
\end{equation}
The equations result in a flux-balance form with the fluxes given by:
\begin{align}
  \label{eq:hydro_fluxes_sphere}
  \vec{f}^{(r)} &\equiv \alpha\sqrt{\gamma^s}\left\{
    D\left(v^r-\frac{\beta^r}{\alpha}\right), \right.\\
  & S_r\left(v^r-\frac{\beta^r}{\alpha}\right) + p, \nonumber\\
  & \left.\tau\left(v^r-\frac{\beta^r}{\alpha}\right)+pv^r\right\} \, \nonumber
\end{align}
and source terms by:
\begin{widetext}
\begin{align}
  \vec{s} &= \left\{ -\frac{2}{r} f^{(r)}_D \right. \ ,  \\
  & - \frac{2}{r}f^{(r)}_{S_r}
  + \alpha\sqrt{\gamma^s} \left(
    T^{00}\left[\half(\beta^r)^2 \partial_r \gamma_{rr} - \alpha \partial_r\alpha \right]
    +  T^{0r}\beta^r \partial_r\gamma_{rr} + T^{0}_r \partial_r \beta^r  
    + \half T^{rr}\partial_r \gamma_{rr} + \frac{p}{\gamma_T} \partial_r \gamma_{T}
  \right) \ , \nonumber \\
  & \left. -\frac{2}{r}f^{(r)}_{\tau}
    + \alpha\sqrt{\gamma^s} \left( T^{00}\left((\beta^r)^2 K_{rr} - \beta^r \partial_r \alpha\right)
      + T^{0r}(2\beta^r K_{rr} - \partial_r \alpha) + T^{rr}K_{rr} +
      \frac{2pK_T}{\gamma_T} \right) 
  \right\} \ .\nonumber
\end{align}
\end{widetext}
Note that, as in the field equations, all the source terms are 
regular, due to the regularity conditions at the origin, since 
$1/r$ terms appear always together with a vector quantity. 
Note also that the system is slightly different from a previous 
formulation, see e.g. Ref.~\cite{Gourgoulhon:1991,Romero:1995cn}, 
and in our case the characteristic speeds are the same as 
the 1D Cartesian one given in \cite{Banyuls:1997}, once
the formal replacements $\gamma\mapsto\gamma^s$, 
$v^x\mapsto v^r$ and $\gamma^{xx}\mapsto\gamma^{rr}$ are applied.

\section{Implementation}
\label{sec:Numerics}

\begin{table}[t]
  \caption{\label{tab:settings} Settings used in the simulations. 
    Note that dimensionless units $c=G=M_\odot=M_{\rm bh}=1$ are employed.}
  \begin{ruledtabular}
    \begin{tabular}{ccccc}
      Test       & 
      Resolution, $N$& 
      Grid       & 
      Atmosphere & 
      Time\\
      & 
      \small{$\{low,med,hig\}$} & 
      \small{$r_{\rm out}$} & 
      \small{$f_{\rm thr},\ f_{\rm lev}$} & 
      \small{$t_{\rm end},\ c_{\rm cfl}$}\\
      \hline
      Flat & 
      $\{1000,2000,4000\}$ &    
      100 & 
      $--,--$ &
      70, 0.5\\
      Puncture    & 
      $\{2000,4000,8000\}$ &     
      200 & 
      $--,--$ & 
      400, 0.5\\
      Stable Star & 
      $\{100,200,400\}$ & 
      20 & 
      $10^{-8},\ 10^{-12}$ & 
      8000, 0.5\\
      Migration   & 
      $\{300,600,1200\}$ & 
      40    & 
      $10^{-6},\ 10^{-9}$ & 
      10000, 0.25\\
      Collapse    & 
      $\{400,800,1600\}$ & 
      20     & 
      $10^{-7},\ 10^{-12}$ & 
      300, 0.5\\
    \end{tabular}
  \end{ruledtabular}
\end{table}

Matter and metric fields are integrated forward in time together 
using the Method of Lines (MoL) based on Runge-Kutta (RK) 
integrators. Both TVD 3rd-order and 4th-order RK, see 
e.g.~\cite{Shu:1988}, were used for the computations subjected to the  
Courant condition for the timestep $\Delta t = c_{\rm cfl} \Delta r$.

The RHS of the equations are discretized on a staggered  
grid in $r\in(0, r_{\rm out}]$ with uniform spacing $\Delta r = r_{\rm out}/N$.
Spatial derivatives of the geometric fields are 
discretized using centered 4th-order finite differences, and 
the 4th-order Kreiss-Oliger dissipation operator is always used in 
simulations, with $\sigma=0.007$. The analysis
of~\cite{Calabrese:2005ft} demonstrates that the standard
discretization of the Z4 formulation in terms of the ADM variables, 
coupled to the Bona-Masso gauge and vanishing shift, will result in an
unstable scheme if artificial dissipation is not used. However It is not
immediately clear whether or not that analysis holds in the
conformally decomposed Z4c case. All the relevant parameters used are 
listed in Tab.~\ref{tab:settings}.

We use two different boundary conditions on the metric variables;
Sommerfeld and maximally dissipative. With BSSN we always use 
Sommerfeld conditions like those typically used in 3D 
implementations. At the boundary each variable is modeled by 
flat space plus a perturbation. One may then derive the boundary 
conditions 
\begin{align}
\p_tf= -vf_{,r}-\frac{v}{r}(f-f_0),
\end{align}
where $f_0$ is the background solution of the field. For Z4 we also 
use maximally dissipative boundary conditions. To construct the 
conditions we linearize the spherical equations of motion around 
flat-space and analyze the system in fully second order form. We
construct the incoming characteristic variables and use conditions 
\begin{align}
U_{-}&=g.
\end{align}
The given data $g$ for the boundary conditions is taken from the 
initial data and held constant. The fully second order characteristic 
variables always contain terms like $\p_0f$ for the evolved 
variables, and may be solved for evolution equations for 
$(\alpha,\beta^r,\chi,\tilde{\gamma}_{rr},\tilde{\gamma}_{T})$ at 
the boundary if one additionally adds the $D$ constraint to the 
boundary conditions. We could use the same method to derive constraint 
preserving conditions, but for now we find that the maximally dissipative 
conditions are satisfactory. At typical resolutions we find that the 
incoming constraint violation is of the order $10^{-7}-10^{-8}$ 
(see Fig.~(\ref{fig:H_star})).

Hydrodynamics equations are solved with an High-Resolution-Shock-Capturing
(HRSC) method based on cell-center discretization and on the
Local-Lax-Friedrichs (LLF) central scheme for the fluxes.
The method is described in \cite{DelZanna:2002qr,DelZanna:2007pk} and 
we refer the reader to these
references for all of the details. The assessment of the LLF flux in 
case of neutron
star evolutions in full GR is presented in \cite{Shibata:2005jv}.
Different reconstruction schemes were tested, overall: the linear TVD based on VanLeer Monotonized
Centered limiter (MC2), the Convex-Essentially-Non-Oscillatory 3rd-order method (CENO3)
and the Piecewise Parabolic Method (PPM). No relevant differences were found
and the data presented were obtained with the CENO3 reconstruction.
The metric fields are reconstructed with either simple averages or \emph{upwinded}
Lagrangian quartic reconstruction.

The vacuum treatment is done with the use of an artificial atmosphere as described
in \cite{Dimmelmeier:2002bk}. When, during the recovery of primitive
variables, a point finish below a certain threshold $\rho_{\rm thr}=f_{\rm thr}\max\left(\rho(t=0)\right)$ all the
variables are set to the atmosphere level $\rho_{\rm lev}=f_{\rm lev}\max\left(\rho(t=0)\right) <\rho_{\rm thr}$.
Typical values are listed in Tab.~\ref{tab:settings}.

In the collapse simulations matter fields were set to the atmosphere value once
the horizon is formed and if $\alpha < \alpha_{\rm hydro-ex}=0.02$. 
This is necessary to avoid numerical problems with the first points close to the origin.

The code has been extensively tested for different problems, especially the HRSC scheme.
As expected, 4th order accuracy is reached in vacuum, 
while matter simulations are accurate at 2nd order 
(except for situation with shocks). 

Stellar initial data are computed with a 2-domain pseudo-spectral method 
to solve the equations of Ref.~\cite{Bonazzola:1993} in isotropic 
coordinates, in spherical symmetry. A $\Gamma=2$ polytropic EoS:
\begin{equation}
  \label{eq:eos_poly}
  P(\rho) = (\Gamma-1)\rho\epsilon \ \  
  \mbox{with} \ \ 
\epsilon = K \frac{\rho^{\Gamma-1}}{\Gamma-1}
\end{equation}
is used with $K=100$ which is compatible with the ideal gas EoS used for the evolution.
Chebyshev-Gauss-Lobatto grids with 64 collocation points per domain are
sufficient for having the equations satisfied to machine accuracy. 



\end{document}